\documentclass[%
 reprint,
 amsmath,amssymb,
 aps, pra,
 longbibliography,
 twocolumn,
 superscriptaddress
]{revtex4-1}

\usepackage{graphicx}
\usepackage{dcolumn}
\usepackage{bm}
\usepackage{siunitx}
\usepackage{nicefrac}
\usepackage{mathrsfs}
\usepackage{dsfont}
\usepackage{lipsum}
\usepackage{multirow}
\usepackage{makecell}

\usepackage[plainpages=false,pdfpagelabels,colorlinks=true,linkcolor=blue,urlcolor=blue,citecolor=blue]{hyperref}

\begin{document}
	
\renewcommand{\vec}[1]{\mathbf{#1}}
\title{Topological Floquet engineering using two frequencies in two dimensions}

\author{Yixiao Wang}
\author{Anne-Sophie Walter}
\affiliation{Institute for Quantum Electronics, ETH Zurich, 8093 Zurich, Switzerland}
\author{Gregor Jotzu}
\affiliation{Max Planck Institute for the Structure and Dynamics of Matter, 22761 Hamburg, Germany}
\author{Konrad Viebahn}
\email{viebahnk@phys.ethz.ch}
\affiliation{Institute for Quantum Electronics, ETH Zurich, 8093 Zurich, Switzerland}

\date{\today}

\begin{abstract}
Using two-frequency driving in two dimensions opens up new possibilites for Floquet engineering, which range from controlling specific symmetries to tuning the properties of resonant gaps.
In this work, we study two-band lattice models subject to two-tone Floquet driving and analyse the resulting effective Floquet bandstructures both numerically and analytically.
On the one hand, we extend the methodology of Sandholzer et al. [10.1103/PhysRevResearch.4.013056] from one to two dimensions and find competing topological phases in a simple Bravais lattice when the two resonant drives at $1\omega$ and $2\omega$ interfere.
On the other hand, we explore driving-induced symmetry breaking in the hexagonal lattice, in which the breaking of either inversion or time-reversal symmetry can be tuned independently via the Floquet modulation.
Possible applications of our work include a simpler generation of topological bands for ultracold atoms, and the realisation of non-linear Hall effects as well as Haldane's parity anomaly in inversion-symmetric parent lattices.
\end{abstract}

\maketitle

\section{Introduction}

Topological phenomena emerge naturally for electrons under the effect of strong magnetic fields, exemplified by the quantum Hall effect (QHE)~\cite{klitzing_new_1980}.
In these phenomena, the underlying physical mechanism is the breaking of time-reversal symmetry due to the magnetic field.
Several years after the discovery of the QHE it was noticed that topological insulators can also arise without an ambient magnetic field, such as the anomalous quantum Hall state~\cite{haldane_model_1988} and the quantum spin Hall effect~\cite{kane_quantum_2005}.
Today, anomalous topological phases can be realised by Floquet engineering, that is, periodic driving of a quantum system.
So far, Floquet engineered topological phenomena have been largely limited to single-frequency protocols, applied to optical lattices~\cite{holthaus_floquet_2016,eckardt_colloquium:_2017,cooper_topological_2019,weitenberg_tailoring_2021,aidelsburger_realization_2013,miyake_realizing_2013,jotzu_experimental_2014,tai_microscopy_2017,flaschner_observation_2018,wintersperger_realization_2020},  real materials~\cite{oka_floquet_2019,rudner_band_2020,harper_topology_2020,mciver_light-induced_2020}, photonic and plasmonic waveguides~\cite{ozawa_topological_2019,rechtsman_photonic_2013,fedorova_observation_2020}, and other synthetic systems~\cite{fleury_floquet_2016,nagulu_chip-scale_2022,chen_digital_2021,yang_observation_2022}.
Recently, bichromatic and multi-frequency Floquet engineering has emerged as a powerful strategy to enhance the driving capabilities.
These include time-reversal or spatial symmetry breaking~\cite{schiavoni_phase_2003,struck_tunable_2012,nag_dynamical_2019,stammer_evidence_2020,jimenez-galan_lightwave_2020,heide_optical_2021,neufeld_light-driven_2021,yao_domain-wall_2022,ikeda_floquet_2022,trevisan_bicircular_2022,olin_topological_2022} and interference between different Floquet harmonics~\cite{zhuang_coherent_2013,niu_excitation_2015,grossert_experimental_2016,gorg_realization_2019,viebahn_suppressing_2021,sandholzer_floquet_2022,castro_floquet_2022}.
The combination of both two-path interference and time-reversal symmetry breaking has led to the proposal~\cite{kang_topological_2020} and experimental demonstration~\cite{minguzzi_topological_2022} of a topological pump in  one-dimensional lattice.
To our knowledge, all experimental results in the multi-frequency regime have been limited to one-dimensional driving patterns, motivating the need to find novel strategies for two-dimensional driving.

In this work, we use the idea of two-tone Floquet engineering to generate novel topological bandstructures and topological phase transitions from simple parent lattices in two dimensions.
On the one hand, we consider resonant driving for the lowest two bands of a triangular lattice.
Although this model only features a single potential minimum  per unit cell, we are able to drive topological transitions by tuning the amplitude and relative phase between $1\omega$ and $2\omega$ drives.
The simplicity of this scheme could open up new pathways for realising strongly correlated topological insulators, which have remained out of reach in existing methods.
On the other hand, we apply two-frequency driving to an inversion-symmetric (graphene-like) hexagonal lattice.
This driving scheme enables the selective breaking of inversion symmetry while maintaining time-reversal symmetry, giving access to a new class of Floquet Hamiltoninans.
Thus, we find that an external breaking of inversion symmetry is not necessary to drive the celebrated parity anomaly in the Haldane model~\cite{haldane_model_1988}.
The two-frequency approach generically applies to Floquet-driven systems in lattices, ranging from condensed matter to synthetic quantum matter.

The remainder of this paper is structured as follows.
The relevant two-tone driving waveforms, and possible experimental implementations, are introduced in section \ref{sec:12}.
Afterwards, we show how resonant $1\omega$--$2\omega$ driving in a triangular lattice leads to topological phase transitions (section \ref{sec:triangular}). Section \ref{sec:hexagonal} discusses the effective breaking of time-reversal and inversion symmetry under off-resonant driving in a hexagonal lattice.

\begin{figure}[htbp]
	\begin{center}
		\includegraphics[width = 0.48\textwidth]{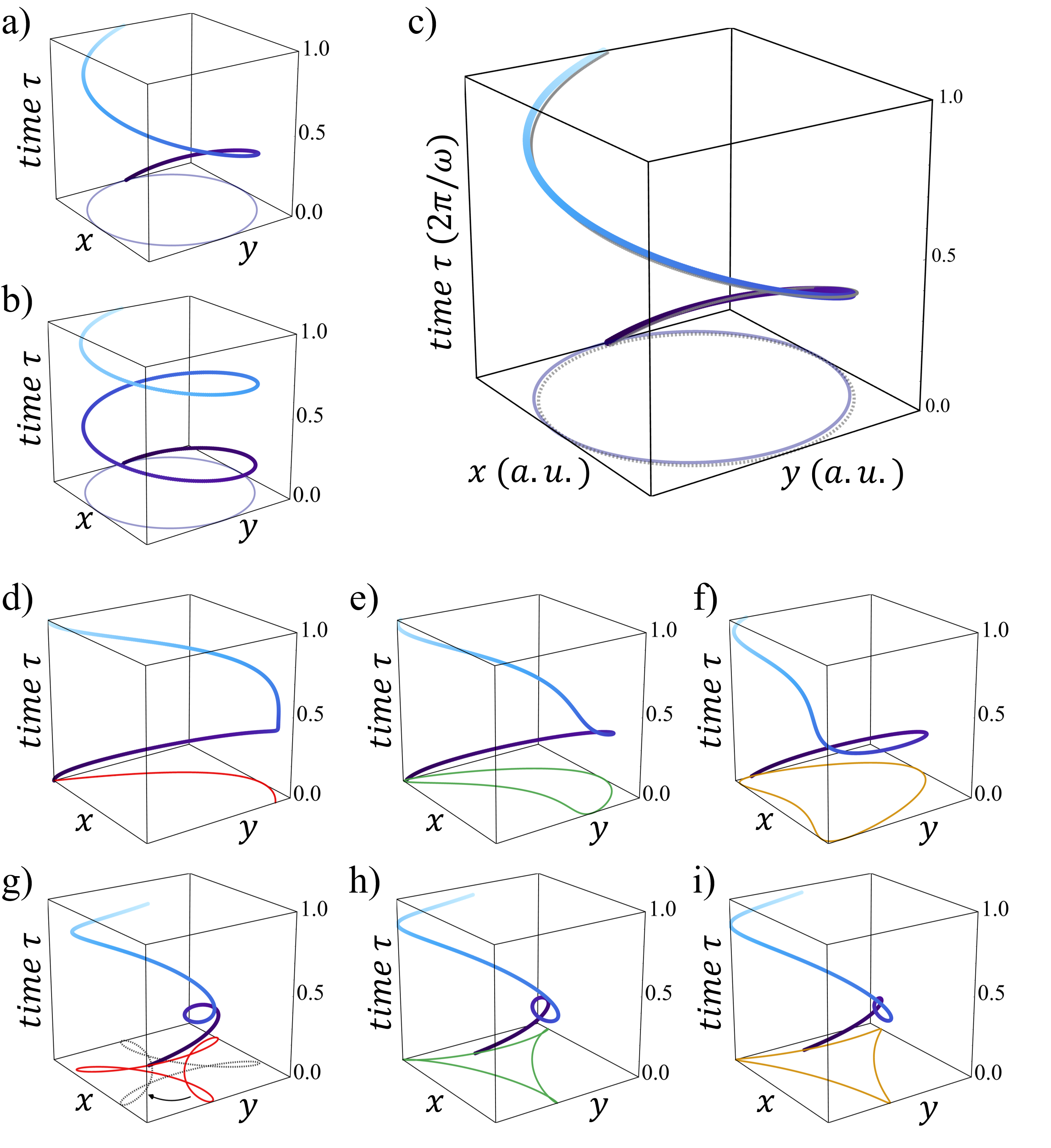}
	\caption{
	Single- and two-frequency Lissajous curves for two-dimensional driving.
 The plots show the real-space modulation figures $\textbf{r}_m(\tau)$ (Eq.~\ref{eqn:shake}). These shapes are topologically equivalent to those of the time-dependent force $\textbf{F}(\tau)$, since the two are related to one another by taking the second derivative (Eq.~\ref{eqn:force}).
   (a-c) Circular strong driving at $1\omega$ (a), circular weak driving $2\omega$ (b), or both drives combined (c) can be used to resonantly address a higher-band transition. The thin line in (c) represents the dominant $1\omega$ waveform, the same as (a).
   (d-f) Superposition of an elliptical driving field at $1\omega$ and a linear field at $2\omega$ with relative phases $\varphi_{y1} = 0$ (d), $0.2\pi$ (e), and $0.5\pi$ (f).
   (g-i) Superposition of two circular drives with relative phase $\varphi_{x2} = 0.9\pi$, $\varphi_{y1} = 0.5\pi$, $\varphi_{y2} = 0.4\pi$, driving strength $K_{x2} = K_{y2} = 0.7$, and $K_{x1} = K_{y1} = 0.5$ (g), $0.65$ (h), and $0.8$ (i).}
	\label{fig:1}
	\end{center}
\end{figure}

\section{$1\omega$--$2\omega$ Floquet driving in two dimensions}\label{sec:12}

Two-frequency driving can be applied, in princple, to any physical system which has been employed for Floquet engineering so far~\cite{oka_floquet_2019}.
As a generic starting point, we consider a particle confined to a static potential $V_{\text{lat}}(r)$, and subject to an oscillating force 
\begin{equation}
    \label{eqn:Hcm}
    \hat{H}_{\text{cm}}(\tau)=\frac{\hat{p}^2}{2m}+V_{\text{lat}}(r)-\textbf{F}(\tau)\cdot\hat{r}~,
\end{equation}
with
\begin{equation}\label{eqn:force}
    \textbf{F}(\tau)\equiv e\textbf{E}(
\tau)=-m\Ddot{\textbf{r}}_m(\tau)=-\dot{\textbf{A}}(\tau)~.
\end{equation}
In solid state systems, the driving force $\vec{F}(\tau)$ can directly correspond to the oscillating field $\textbf{E}(\tau)$ of a laser coupling to free electrons~\cite{wang_observation_2013,mciver_light-induced_2020}.
Equivalently, the application of periodic driving can be seen as `minimal coupling' using the vector potential as $q \rightarrow q - \textbf{A}(\tau)/\hbar$.
Furthermore, collective modes such as phonons~\cite{disa_engineering_2021} or magnons~\cite{vinas_bostrom_light-induced_2020} can also be involved.
In waveguide-based synthetic lattices, transverse movements of the waveguides provide an effective inertial force~\cite{rechtsman_photonic_2013}.
For optical lattices, the driving force can be provided by an oscillating optical or magnetic gradient~\cite{jotzu_creating_2015}.
Alternatively, an inertial force can result from a periodic displacement of the entire lattice potential, an approach known as `lattice shaking'~\cite{holthaus_floquet_2016,eckardt_colloquium:_2017}.
In this case, the Hamiltonian in the lab frame is
\begin{equation}\label{eqn:Hreal}
    \begin{aligned}
        \hat{H}_{\text{lab}}(\tau)&=\frac{\hat{p}^2}{2m}+V_{\text{lat}}\left[r-\textbf{r}_m(\tau)\right]~,
    \end{aligned}
\end{equation}
related to Eq.~\ref{eqn:force} by two unitary transformations~\cite{dalibard_reseaux_2013}.
In summary, the derivations in this work directly carry over from engineered quantum platforms, such as ultracold atoms in optical lattices, to condensed matter systems.

We define the two-dimensional driving pattern $\textbf{r}_m(\tau)$ as
\begin{eqnarray}
    \label{eqn:shake}
        x_m(\tau)= & A_{x1}\cos(\omega\tau) & +  A_{x2}\cos(2\omega\tau+\varphi_{x2})~\\
        \nonumber y_m(\tau)= & A_{y1}\cos(\omega\tau+\varphi_{y1}) & + A_{y2}\cos(2\omega\tau+\varphi_{y2})~.
\end{eqnarray}
These two-tone driving waveforms are visualised as Lissajous curves in Fig.~\ref{fig:1} for the relevant choices of driving parameters (phases and amplitudes) used in this work.
The amplitude of the vector potential is given by the dimensionless driving strength $K_{\alpha\beta}=m\omega_{\beta} A_{\alpha\beta} a/\hbar$, where $a$ is the lattice spacing and $A_{\alpha\beta}$ is the real-space amplitude.
The index $\alpha\in \{x,y\}$ denotes the direction, whereas $\beta \in \{1,2\}$ indicates the frequency.

\textit{Resonant circular $1\omega$--$2\omega$ driving.} The combination of a strong circular $1\omega$ drive (Fig.~\ref{fig:1}a, $\varphi_{y1} = \pi/2$) with a weak circular $2\omega$ drive (Fig.~\ref{fig:1}b, $\varphi_{x2} = 0$, $\varphi_{y2} = \pi/2$) leads to a modulated circular pattern (Fig.~\ref{fig:1}c).
Due to two-path interference when addressing the $s$--$p$ resonance of the triangular lattice, this driving pattern can be used to drive topological transitions in two dimensions, similar to the topological pump of ref.~\cite{minguzzi_topological_2022}.

\textit{Off-resonant $1\omega$--$2\omega$ driving to access spatial and temporal symmetries.} Alternatively, an elliptical driving field at $1\omega$ and a linearly polarised laser at $2\omega$ ($A_{x2} = 0, \varphi_{y2} = 0$) leads to a competition between the breaking of inversion symmetry and time-reversal symmetry.
For instance, the choice $\varphi_{y1} = 0$ results in a `boomerang' pattern (Fig.~\ref{fig:1}d), which breaks inversion symmetry but not time-reversal symmetry.
Conversely, the case of $\varphi_{y1} = \pi/2$ (Fig.~\ref{fig:1}f) gives a rounded `kite' shape in which the breaking of time-reversal symmetry dominates over the breaking of inversion symmetry.
Tuning the phase $\varphi_{y1}$ thus allows to interpolate between the two symmetry breaking situations (Fig.~\ref{fig:1}e), thereby accessing novel topological regimes.

Instead of combining linear and elliptical drives, the competition between inversion-breaking and time-reversal-breaking can be achieved by superimposing two circular drives of opposite helicity (Fig.~\ref{fig:1}g-i).
The way in which inversion symmetry is broken in this case depends on the relative orientation of the driving pattern to the lattice geometry. This can be tuned by increasing $\varphi_{x2}$ and $\varphi_{y2}$ by the same amount (i.e.~by delaying the two frequency drives with respect to each other), which rotates the driving pattern around the origin (Fig.~\ref{fig:1}g, see also refs.~\cite{jimenez-galan_lightwave_2020,trevisan_bicircular_2022,ikeda_floquet_2022}).
Driving a topological transition via changing only a relative phase can be experimentally advantageous in condensed matter systems where changing polarisation or amplitude can lead to spurious effects. 

\textit{Experimental implementation.} 
In this work we focus on a triangular and a hexagonal lattice. A corresponding optical lattice potential, and the resulting tight-binding parameters, can be found in Appendix~\ref{app:optical-lattice}.
In waveguide-based systems, the lattice geometry can generally be chosen freely~\cite{rechtsman_photonic_2013}.
The time-dependence of the driving term is encoded in an additional spatial coordinate, which can accommodate multi-frequency driving. 
Finally, in solid state systems, the laser field corresponding to the driving term at $2\omega$ can be generated via second-harmonic generation, ensuring that it is phase-stable with respect to the $1\omega$ term. The relative phase between the two can then be tuned using a dispersive material of varying thickness.

\begin{figure}[t]
	\begin{center}
		\includegraphics[width = 0.48\textwidth]{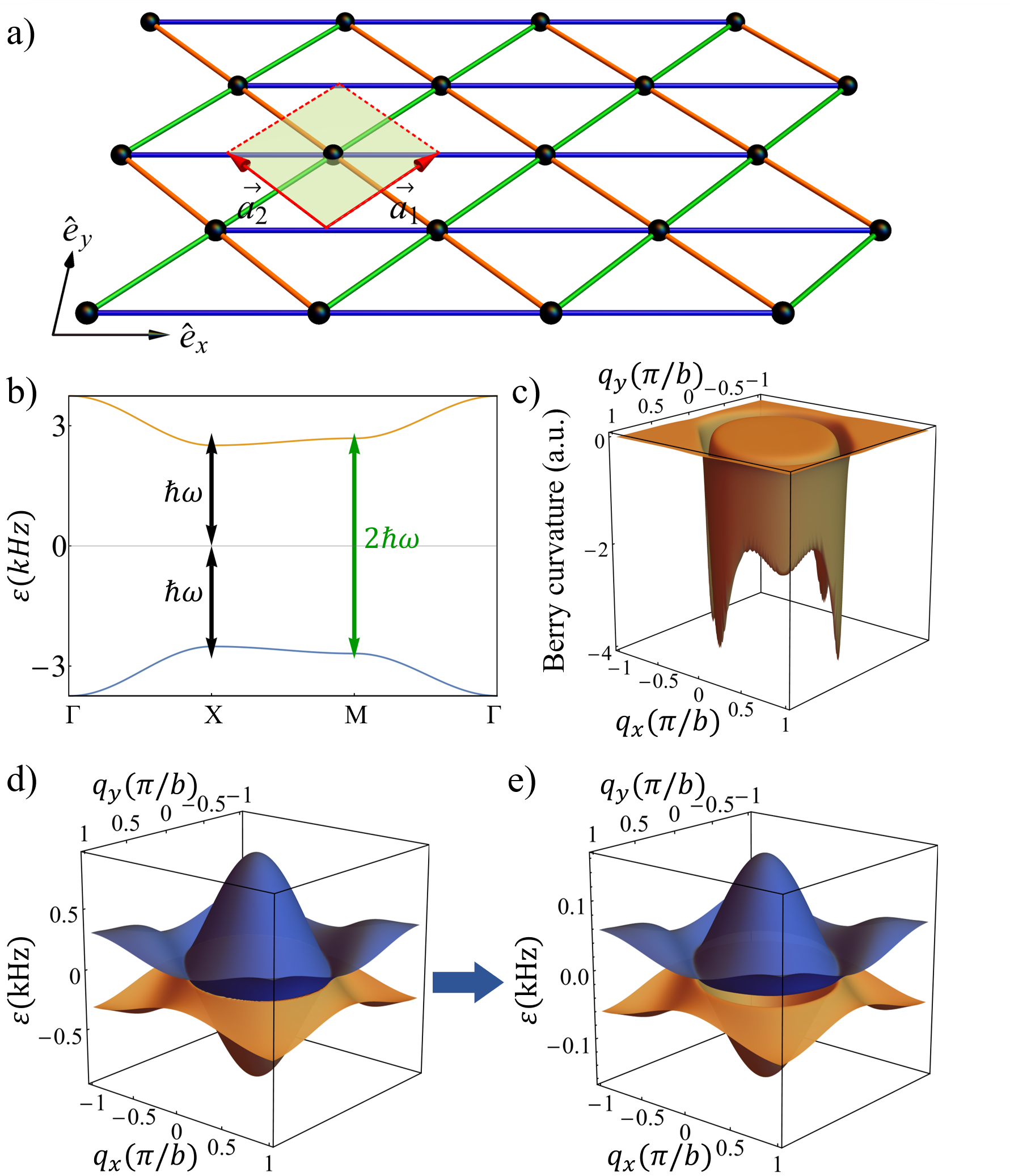}
		\caption{
  Resonant band coupling in the triangular lattice (a), with relevant parameters $t_1=-0.05$, $t_2=-0.07$, $t_3=-0.07$, $\varepsilon=-0.66$ in units of $E_{rec}$ and dimensionless inter-band coupling $\eta_{sp}=0.19$. Resonant one-photon or two-photon processes lead to the coupling of lower and higher bands, corresponding to the green and black arrows in (b), respectively. Floquet driving lifts the degeneracy of the static band structure in the rotating frame (c) and opens a band gap (d). The ring-shape minima of the band structure leads to a ring-shape Berry curvature (e). (f) shows a trivial spatially localized Berry curvature under two-frequency driving, which features a strong Berry curvature dipole.
		}
		\label{fig:2}
	\end{center}
\end{figure}

\textit{Tight-binding Hamiltonians for two-dimensional Floquet engineering.} 
By expanding the original Hamiltonian (Eq.~\ref{eqn:Hcm}) in a Wannier basis and transforming it to quasimomentum space, we get the following expression
\begin{equation}
\label{eqn:Hq}
    \begin{aligned}
        &\sum_{n}\biggl\{\varepsilon_{n}\hat{a}_n^{\dagger}\hat{a}_n-\sum_{i=1}^{Z}\left[t_{n, i}\left(e^{i(\theta_i(\tau)+\textbf{q}\cdot\textbf{b}_i)}\hat{a}_n^{\dagger}\hat{a}_n+h.c.\right)\right]
        \\&-\textbf{F}(\tau)\cdot\hat{r}\sum_{n'}\eta_{nn'}\hat{a}_n^{\dagger}\hat{a}_{n'}\biggl\}~.
    \end{aligned}
\end{equation}
Here, $\varepsilon_n$, $t_{n,i}$, $\eta_{nn'}$ are the band centre energies,  nearest-neighbour tunnelling matrix elements, and interband coupling elements, respectively, for bands $n$, $n'$ (Appendix~\ref{app:wannier}).
$\textbf{b}_i=j\textbf{a}_1+k\textbf{a}_2$ are the nearest-neighbour tunneling vectors and $i$ indexes the in-equivalent neighbours up to the coordination number $Z$; $\textbf{a}_1$, $\textbf{a}_2$ correspond to the primitive translation vectors in two dimensions.
The time-periodic Peierls phases are $\theta_i(\tau)=\frac{m}{\hbar} \textbf{b}_i\cdot\dot{\textbf{r}}_m(\tau)$.
In second quantisation, $\hat{a}_{n}$ denotes the annihilation operator in band $n$. 
The first line of Eq.~\ref{eqn:Hq} describes static and time-dependent contributions to intra-band processes whereas the second line is the inter-band coupling that arises from periodically forcing the system.
The coupling elements $\eta_{nn'}$ can be understood as dipole matrix elements of Bloch states~\cite{sandholzer_floquet_2022}.
Consequently, the first line is diagonal in quasimomentum and band index $n$, whereas the second line contributes off-diagonal elements to the Hamiltonian.

We use the Hamiltonian in Eq.~\ref{eqn:Hq} as the starting point for the Floquet analysis.
If the periodic drive resonantly addresses the gap between $s$ and $p$ bands, we additionally employ a unitary transformation to `rotate away' the energy difference (sec.~\ref{sec:triangular}).
In order to evaluate the effective Floquet Hamiltonian, we employ two complementary methods.
On the one hand, we calculate the effective Hamiltonian analytically in inverse powers of the driving frequency $\omega$ using the well-known high-frequency expansion (HFE)~\cite{rahav_effective_2003,goldman_periodically_2014,eckardt_high-frequency_2015,bukov_universal_2015,mikami_brillouin-wigner_2016}.
Since we work in the weak driving regime the HFE remains valid.
On the other hand, we can numerically obtain the effective Hamiltonian via the Trotter decomposition~\cite{trotter_product_1959}.
We use this numerical method to validate the results of the analytic calculation and to justify the truncation of the high-frequency expansion a posteriori. 
To characterise the band topologies, the Berry curvature is evaluated via the eigenstates in the discretised two-dimensional Brillouin zone.
The resulting Chern number is the surface integral of the Berry curvature over the closed Brillouin torus~\cite{fukui_chern_2005,asboth_short_2016}.

\section{Resonantly driven triangular lattice}\label{sec:triangular}

The first situation we consider is a simple triangular lattice in which the lowest two bands ($s$ and $p$) are resonantly coupled with two frequencies, inspired by ref.~\cite{kang_topological_2020} and building on previous works using single-frequency driving~\cite{baur_dynamic_2014,zheng_floquet_2014,zhang_shaping_2014}.

In the non-shaken case the static Hamiltonian for a triangular tight-binding model is
\begin{equation}\label{eqn:triangular}
    \hat{H}=\left[\varepsilon+\sum_{i=1}^3 t_i\cos(\textbf{q}\cdot\textbf{b}_i)\right]\sigma_z~.
\end{equation}
The nearest-neighbour tunneling vectors are $\textbf{b}_1=(-1,1)b$, $\textbf{b}_2=(1,0)b$, $\textbf{b}_3=(0,1)b$ and $\textbf{q}=(q_x,q_y)$ in the rotated coordinate $\hat{e}_x'=\frac{\hat{e}_x+\hat{e}_y}{\sqrt{2}}$, $\hat{e}_y'=\frac{-\hat{e}_x+\hat{e}_y}{\sqrt{2}}$ (Fig.~\ref{fig:2}a).
The simplest way to couple $s$ and $p$ bands is to apply single-frequency driving, parametrised by Eq.~\ref{eqn:shake} with $A_{x2} = A_{y2} = 0$ and denoting $\varphi_{y1}$ as $\varphi$.

Starting from Eq.~\ref{eqn:Hq} the time-dependent Hamiltonian can be written as
\begin{equation}
    \begin{aligned}
        \hat{H}(\tau)=&\left\{\varepsilon+\sum_{i=1}^3 t_i\cos\left[\textbf{q}\cdot\textbf{b}_i+\theta_i(\tau)\right]\right\}\sigma_z+\\
        &\left[\eta_{sp}\hbar\omega K_x\cos(\omega\tau)+\eta_{sp}\hbar\omega K_y\cos(\omega\tau+\varphi)\right]\sigma_x,
    \end{aligned}
\end{equation}
where $\varepsilon=(\varepsilon_s-\varepsilon_p)/2=-\Delta\varepsilon/2$ with $\Delta\varepsilon$ being the gap size.
The other parameters are $t_i=t_{s,i}-t_{p,i}$ with $i={1,2,3}$, the Peierls phases are $\theta_i(\tau)=-\frac{1}{\hbar}\int_0^{\tau}\textbf{F}(\tau')\cdot\textbf{b}_i\,d\tau'=-[K_x\sin(\omega\tau)\hat{x}+K_y\sin(\omega\tau+\varphi)\hat{y}]\cdot\textbf{b}_i/b$, and the Pauli matrices are $\sigma_j$ with $j=x, y, z$. The dimensionless driving strengths are denoted by $K_x=m\omega A_{x1} a/\hbar$, $K_y=m\omega A_{y1} a/\hbar$, respectively. The tight-binding parameters, including the tunneling amplitudes, inter-band coupling and band center energies, are obtained from a triangular lattice potential by evaluating the Wannier functions (Appendix~\ref{app:wannier}, Fig.~\ref{fig:2}a).

\begin{figure}[htbp]
	\begin{center}
		\includegraphics[width = 0.5\textwidth]{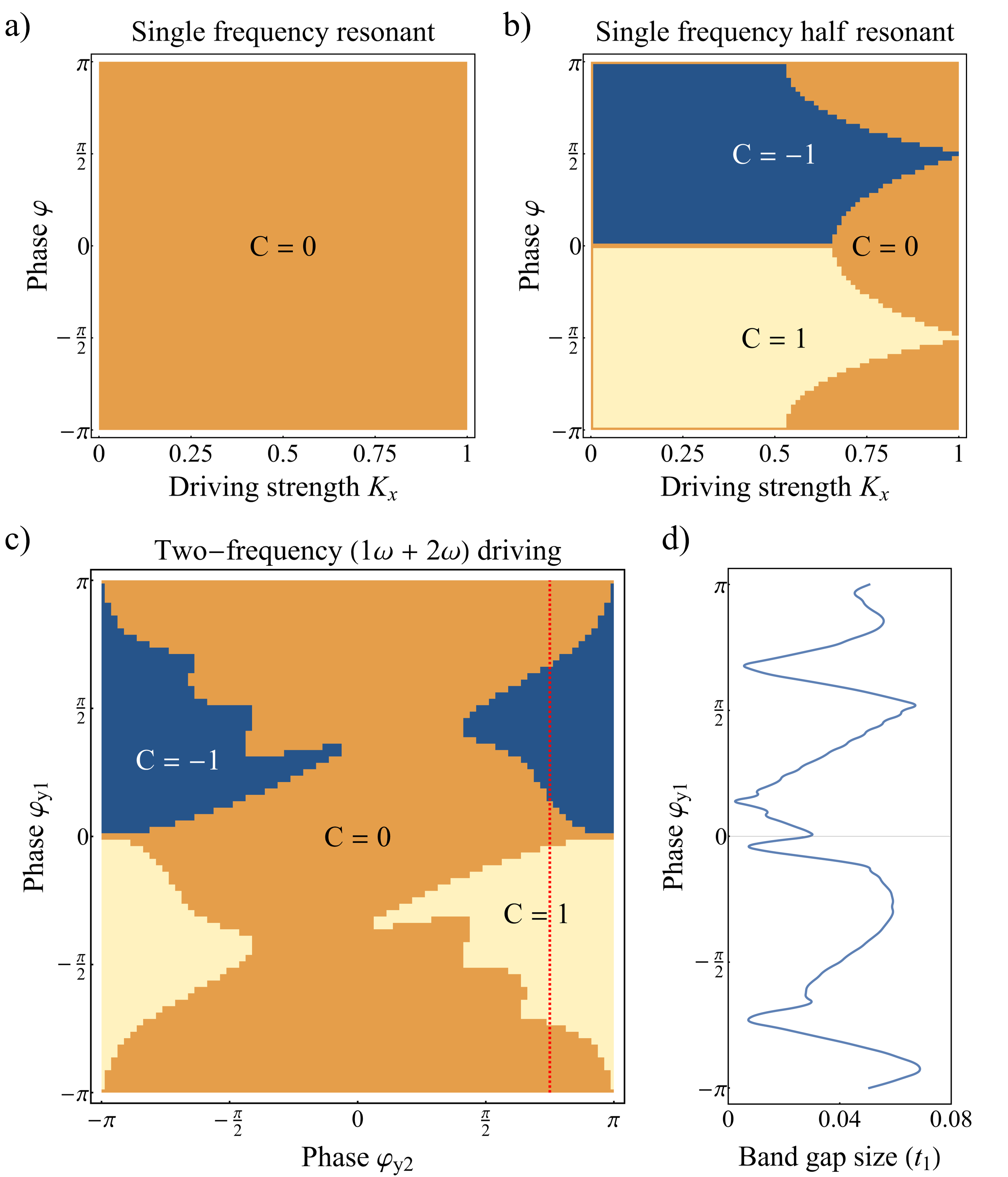}
		\caption{
Phase diagrams of the triangular lattice under resonant driving.
Single-frequency driving on resonance ($\hbar\omega = 25.6 t_1 = 1.28 E_{rec}$, $K_y = 0.5$, $\varepsilon + \hbar\omega/2\simeq 0$) leads to trivial bands (a), whereas driving on half-resonance ($\hbar\omega = 12.8 t_1= 0.64 E_{rec}$, $K_y = 0.5$, $\varepsilon + \hbar\omega\simeq 0$) induces topological bands (b). The trivial regions in (b) are caused by bands being shifted out of resonance due to the $ac$-Stark effect. (c) shows the competition between the two resonant processes induced by the two-frequency driving with $K_{x1} = 0.4, K_{y1} = 0.5, K_{x2} = 0.02, K_{y2} = 0.02$. (d) resulting band gap evaluated for a cut through (c) along the red line.
		}
		\label{fig:3}
	\end{center}
\end{figure}

\textit{Single-frequency circular driving resonant with the gap.}
If the frequency $\omega$ is near-resonant with the gap, corresponding to the dominant one-photon process, we apply the unitary $\mathcal{U}_R(\tau)=\exp(-i\omega\tau\sigma_z/2)$ to eliminate the energy gap between the two bands (Fig.~\ref{fig:2}d).
This gives
\begin{equation}
    \begin{aligned}
    \label{eqn:th1}
        \hat{H}(\tau)=&\left\{\varepsilon+\hbar\omega/2+\sum_{i=1}^3 t_i\cos\left[\textbf{q}\cdot\textbf{b}_i+\theta_i(\tau)\right]\right\}\sigma_z\\&+\left[\eta_{sp}\hbar\omega K_x\cos(\omega\tau)+\eta_{sp}\hbar\omega K_y\cos(\omega\tau+\varphi)\right]\\
        &\times\left[\sigma_x\cos(\omega\tau)-\sigma_y\sin(\omega\tau)\right]/2.
    \end{aligned}
\end{equation}
The terms proportional to the identity matrix can be omitted without affecting the topology of the bands.

By analysing Eq.~\ref{eqn:th1} in the high-frequency expansion (Appendix~\ref{app:HFE}), we find a ring-shaped gap opening (Fig.~\ref{fig:2}e, see also refs.~\cite{wintersperger_realization_2020,bracamontes_realization_2022}).
However, the off-diagonal terms are dominanted by constants.
Physically, this corresponds to on-site couplings between the $s$ and $p$ bands, which yield topologically trivial bands (Fig.~\ref{fig:3}a).

\textit{Single-frequency elliptical driving resonant with half the gap.}
Driving the system at half the gap energy ($\hbar\omega = 12.8 t_1= 0.64 E_{rec}$) leads to neighbouring-site interband couplings which can give rise to topological bands.
Starting from the waveforms in Eq.~\ref{eqn:shake} with $K_{x2} = K_{y2} = 0$ and applying the unitary $\mathcal{U}_R(\tau)=\exp(-i\omega\tau\sigma_z)$ gives
\begin{equation}
    \begin{aligned}
    \label{eqn:th2}
        \hat{H}(\tau)&=
        \left\{\varepsilon+\hbar\omega+\sum_{i=1}^3 t_i\cos\left[\textbf{q}\cdot\textbf{b}_i+\theta_i(\tau)\right]\right\}\sigma_z\\
        &+[\eta_{sp}\hbar\omega K_x\cos(\omega\tau)+\eta_{sp}\hbar\omega K_y\cos(\omega\tau+\varphi)]\\
        &\times[\sigma_x\cos(2\omega\tau)-\sigma_y\sin(2\omega\tau)]/2,
    \end{aligned}
\end{equation}
where $\varepsilon+\hbar\omega\approx0$.
In this case, the only non-zero Fourier components of the drive are $\pm1\omega$ and $\pm3\omega$.
Contrary to the direct one-photon resonance, the inter-band couplings happen via two-photon processes (Fig.~\ref{fig:2}b).
We now find quasimomentum-dependent $\sigma_x$ and $\sigma_y$ terms in the effective Hamiltonian (Appendix~\ref{app:HFE}), leading to topologically non-trivial regions in the phase diagram (Fig.~\ref{fig:3}b, see also ref.~\cite{zhang_shaping_2014}).
Lines at $K_x = 0$ and $\varphi = 0$ have zero extent in the phase diagram of Fig.~\ref{fig:3}b, preventing the observation  of a transition from $C \neq 0$ to $C = 0$ in a realistic experiment.
While large driving amplitudes (beyond $K_x = 0.5$) render the model topologically trivial, this effect is not due to an interference between different processes, but rather due to the $ac$-Stark effect that shifts the bands out of resonance~\cite{holthaus_floquet_2016}.
Combining the above two resonant schemes at $1\omega$ and $2\omega$  introduces genuine topological transitions as function of driving phase, as outlined in the following.

\textit{Two-frequency resonant elliptical driving.} The competition between resonant $1\omega$ and $2\omega$ driving results in an interplay between topological phases with $C = 0$ and $C = \pm 1$.
We now consider the full two-tone waveforms of Eq.~\ref{eqn:shake} for fixed $[K_{x1},K_{x2},K_{y1},K_{y2}]$, and tuneable $\varphi_{y1}$ and $\varphi_{y2}$ ($\varphi_{x2} = 0$).
This Floquet scheme allows to cross topological transitions purely by changing the phase between $1\omega$ and $2\omega$ drives, as is evident from the distinct regions of $C= 0$ and $C \neq 0$ in the phase diagram (Fig.~\ref{fig:3}c).
The values of the induced band gaps are on the order of $0.1$ to $0.2$ tunnelling energies, which can be enlarged by choosing stronger driving amplitudes.

\begin{figure}[b]
	\begin{center}
		\includegraphics[width = 0.5\textwidth]{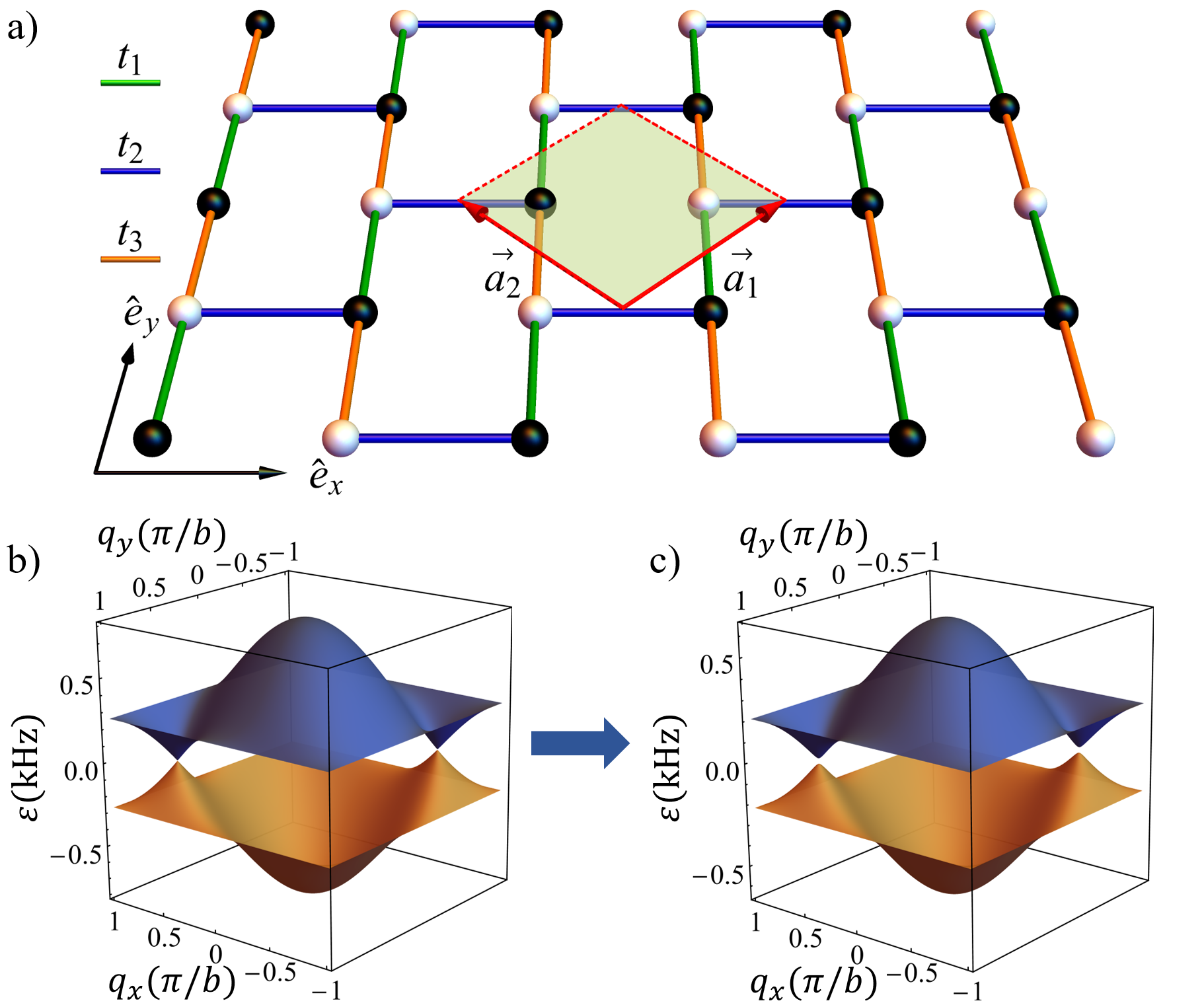}
		\caption{
Off-resonant driving in the hexagonal lattice (a) with tunnelings $t_1 = t_2 = t_3 = 0.06$ in units of $E_{rec}$. The static band structure features two degenerate Dirac points (b). Floquet driving lifts the degeneracy and opens a band gap (c) by effectively breaking symmetries.
		}
		\label{fig:4}
	\end{center}
\end{figure}

\begin{figure*}[t!]
	\begin{center}
		\includegraphics[width = 0.9\textwidth]{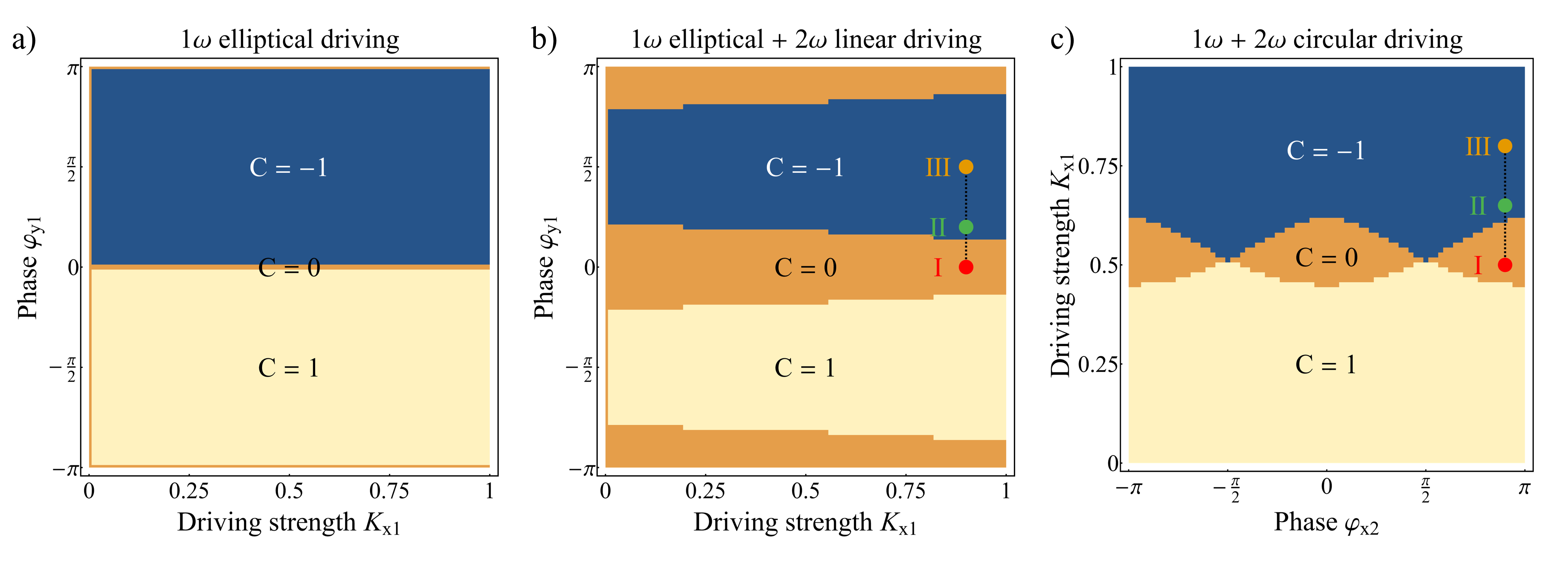}
		\caption{
Phase diagrams of the hexagonal lattice under off-resonant driving ($\hbar\omega = 20.8t_1 = 1.25E_{rec}$). Single-frequency elliptical driving ($K_{y1} = 0.9$, $K_{x2} = K_{y2} = 0$) can break time-reversal symmetry and result in a non-zero Chern number (a). The breaking of the inversion and time-reversal symmetries can be independently controlled by adding a linearly polarized field ($K_{y1} = K_{y2} = 0.9$, $K_{x2} = 0$ and $\varphi_{x2} = \varphi_{y2} = 0$) (b), as well as using two circular fields with opposite polarization ($K_{x1} = K_{y1}$, $K_{x2} = K_{y2} = 0.7$, $\varphi_{y1} = \pi/2$ and $\varphi_{y2} = \varphi_{x2} - \pi/2$) (c). The transition paths I - II - III marked in (b) and (c) correspond to the Lissajous curves in Fig.~\ref{fig:1}d-f and Fig.~\ref{fig:1}g-i, respectively.
		}
		\label{fig:5}
	\end{center}
\end{figure*}

Realising topological band structures in a Bravais lattice, such as the simple triangular lattice considered here, has important implications.
For instance, it simplifies the generation of topological bands in the context of ultracold atoms.
Up to now, two-dimensional topological models in optical lattices have been realised either in bipartite lattices~\cite{jotzu_experimental_2014,flaschner_observation_2018,zhang_shaping_2014}, square lattices with moving superlattices~\cite{aidelsburger_realization_2013,miyake_realizing_2013,tai_microscopy_2017}, or by employing spin-orbit coupling~\cite{sun_highly_2018}.
These implementations rely on the relative phase stability between lattice or Raman laser beams, leading to a significant technological overhead.
An insufficient phase stability may have been an obstacle for realising strongly correlated phases in two-dimensional topological lattices.
The triangular lattice, on the contrary, can be built by simply superimposing three standing waves in the plane with no active stabilisation.
Therefore, the $1\omega$--$2\omega$ Floquet scheme could provide an avenue towards realising correlated topological states of matter.
In addition to its conceptual simplicity, the two-tone driving leads to the appearance of Berry curvature along a ring-shaped gap, which can be asymmetrically distributed (Fig.~\ref{fig:2}f), particularly in the topologically trivial regions of Fig.~\ref{fig:3}c.
This supports a non-zero Berry curvature dipole, enabled by periodic driving, although the underlying lattice is trivial and inversion-symmetric.
This finding is relevant beyond the field of ultracold atoms and it could enable the observation of nonlinear Hall effects which so far have relied on materials with broken inversion symmetry~\cite{du_nonlinear_2021}.

\section{Off-resonantly driven hexagonal lattice}\label{sec:hexagonal}

The second application of the two-frequency driving method considers the hexagonal lattice in brick-wall configuration, shown in Fig.~\ref{fig:4}a.
This potential can be realised in the same setup as the triangular lattice of sec.~\ref{sec:triangular}, but the derivation also applies to $120^{\circ}$ honeycomb lattices~\cite{flaschner_observation_2018} as well as real graphene~\cite{mciver_light-induced_2020}.

We start from the tight-binding Hamiltonian
\begin{equation}
    \begin{aligned}
        \hat{H}_{\bm{q}}(\tau)=&\left\{\sum_{i=1}^3 t_i\cos\left[\textbf{q}\cdot\textbf{b}_i+\theta_i(\tau)\right]\right\}\sigma_x+
        \\&\left\{\sum_{i=1}^3 t_i\sin\left[\textbf{q}\cdot\textbf{b}_i+\theta_i(\tau)\right]\right\}\sigma_y,
    \end{aligned}
\end{equation}
The nearest-neighbour tunneling vectors are $\textbf{b}_1=(0,1)b$, $\textbf{b}_2=(-1,1)b$ and $\textbf{b}_3=(-1,0)b$ in the rotated coordinate system $\{\hat{e}_x',\hat{e}_y'\}$ (Fig.~\ref{fig:4}a). 
The waveforms of Eq.~\ref{eqn:shake} result in the following Peierls phases:
\begin{equation}
    \begin{aligned}
        &\theta_1(\tau)=-K_{y1}\sin(\omega\tau+\varphi_{y1})-K_{y2}\sin(2\omega\tau+\varphi_{y2}) = -\theta_3(\tau),
        \\&\theta_2(\tau)=K_{x1}\sin(\omega\tau)+K_{x2}\sin(2\omega\tau+\varphi_{x2}),
    \end{aligned}
\end{equation}
with the driving strengths being $K_{x1}=m\omega A_{x1} a/\hbar$, $K_{x2}=2m\omega A_{x2} a/\hbar$, $K_{y1}=m\omega A_{y1} a/\hbar$, and $K_{y2}=2m\omega A_{y2} a/\hbar$, as before.
Since we do not consider inter-banding couplings here, there is no need to apply unitaries and the high-frequency expansion can be directly conducted. 

\textit{Single-frequency elliptical driving.}
The choice $K_{x2} = K_{y2} = 0$, $K_{x1}>0$ corresponds to elliptical driving, giving rise to topological or trivial bands.
For any $K_{y1} > 0$ and $\varphi_{y1} \mod \pi \neq 0$ we recover the well-known topological gap opening at both Dirac points due to time-reversal symmetry breaking (Fig.~\ref{fig:4}b-c)~\cite{oka_photovoltaic_2009}.
The limits of $K_{y1} = 0$ or $\varphi_{y1} \mod \pi = 0$ correspond to linearly polarised driving, leaving time-reversal symmetry intact and both Dirac points remain closed. However, the trivial region with $C = 0$ is only a `line' of zero extent in the phase diagram (Fig.~\ref{fig:5}a, as before in Fig.~\ref{fig:3}b).
Usually, a manifestly broken inversion symmetry of the underlying lattice is required to cross a topological phase transition and induce Haldane's `parity anomaly', thereby extending the trivial `lines' in the phase diagram to finite regions~\cite{haldane_model_1988, jotzu_experimental_2014}.
In the following, we demonstrate that inversion symmetry can be broken by employing the two-tone Floquet drive only.
Therefore, a topological phase transition between trivial and non-trivial regions can be achieved purely via driving.

\textit{Two-frequency driving.}
Starting from the above case of single-frequency elliptical driving, we add linearly polarized `light' by setting $K_{x2} = 0$ while $K_{y2} \neq 0$.
The driving field further breaks inversion symmetry and the competition between the breaking of inversion and time-reversal symmetry can be observed in Fig.~\ref{fig:5}b-c.
The topological phase transition can simply be induced by tuning the phase along the marked path, corresponding to the Lissajous curves in Fig.~\ref{fig:1}d-f.

Alternatively, constraining $K_{x1} = K_{y1}$, $K_{x2} = K_{y2}$, $\varphi_{y1} = \pi/2$, and $\varphi_{y2} = \varphi_{x2} - \pi/2$ in Eq.~\ref{eqn:shake}, the $1\omega-2\omega$ driving scheme corresponds to two counter-rotating, circularly polarized fields.
Clearly, when one circular driving field is much larger than the other, the resulting Chern number will correspond to the helicity of the dominant field.
However, when the amplitude of one circular field approaches the other one, the Lissajous curve becomes a `cloverleaf' pattern and the breaking of time-reversal symmetry is suppressed (Fig.~\ref{fig:1}g-i).
This competition allows to interpolate between topologically trivial and non-trivial settings (with either positive or negative mass terms) purely by changing driving phases.
The resulting gap sizes are similar to the known results in the Haldane model~\cite{jotzu_experimental_2014}.
Interestingly, we found that a pure Dirac Hamiltonian subject to two-frequency driving does not show a corresponding gap opening in the case of broken inversion symmetry.
While time-reversal symmetry breaking via circular driving can be captured within the approximate Dirac Hamiltonian~\cite{mciver_light-induced_2020}, the inversion symmetry breaking relies on the presence of a lattice potential.
Mathematically, this can be seen from the fact that the driven Dirac Hamiltonian remains linear and its Fourier components at $\omega$ and $2\omega$ will not mix in the high-frequency expansion.
Conceptually, a specific driving pattern favours a gap opening specific to one sublattice of the honeycomb, which does not apply to the Dirac Hamiltonian.

We also studied the behaviour of inversion symmetry breaking beyond the validity of the high-frequency expansion, using numerically calculated band structures.
Here, the gap opening scales roughly as $1/\omega^4$ which becomes dominant in the low-frequency regime.
Therefore, the $1\omega$--$2\omega$ scheme could be particularly interesting for phase-controlled topological transitions in laser-driven graphene, which could be readily implemented in setups such as used in ref.~\cite{mciver_light-induced_2020}.


\section{Conclusion}\label{sec:conclusion}

We have shown that two-frequency Floquet driving in two dimensions gives rise to rich topological phase diagrams, both for resonant and off-resonant modulation.
We identify the relative phase between the $1\omega$ and $2\omega$ drives as an important control parameter over the symmetry properties and the resulting topologies in the  effective Floquet Hamiltonians.
In many experiments, phase-only control is advantageous over other parameters, such as frequencies or amplitudes.
For periodically driven real materials, such as graphene~\cite{mciver_light-induced_2020}, incoherent excitations and electron relaxation can strongly affect the measurements, potentially obscuring any topological response.
Incoherent effects often depend on the polarisation of the incident light, especially close to the contacts of a sample.
Several of the driving schemes proposed in this work, particularly the case of counter-rotating, circular drives, depend only on the relative phase between the light fields.
Therefore, we believe that our approach will lead to novel applications in manipulation and detection of topological phenomena.
By combining the schemes considered here with spin-dependent Floquet engineering~\cite{jotzu_creating_2015}, it may become possible to access the entire `periodic table' of topological insulators~\cite{kitaev_periodic_2009}.

\section*{Acknowledgments}
We acknowledge funding by the Swiss National Science Foundation (Grants No.~182650 and NCCR-QSIT) and European Research Council advanced grant TransQ (Grant No.~742579).


%

\clearpage

\onecolumngrid
\appendix

\section{Optical lattice potential}\label{app:optical-lattice}

Lattices of various geometric configurations can be readily realised in the setup of ref.~\cite{uehlinger_artificial_2013}, that is,
\begin{equation}
    \begin{aligned}
    \label{eqn:potential}
        &V_{\text{lat}}(x, y)=\\
        &-V_{\overline{x}}\cos(k_Lx+\theta/2)^2-V_x\cos(k_Lx)^2-V_y\cos(k_Ly)^2\\&-2\sqrt{V_xV_y}\cos(k_Lx)\cos(k_Ly)\cos(\phi), 
    \end{aligned}
\end{equation}
where $k_L=2\pi/\lambda$.
The spacing between neighbouring minima of a one-dimensional standing wave is $a=\lambda/2$, the recoil energy is $E_{rec}=(\hbar k_L)^2/(2m)$, the wavelength $\lambda$ is \SI{1064}{nm}, and the atoms are potassium-40.
The following choice of parameters: $V_{\overline{x}}=0.6$, $V_x=0.3$, $V_y=2.8$ all in units of $E_{rec}$, $\theta=\pi$, and $\phi=0$ yields a triangular lattice. The hexagonal lattice can be realized by setting $V_{\overline{x}}=12$, $V_x=0.8$ and $V_y=4.65$.
Compared to the tight-binding lattices, we have the length of the unit cells $b=\sqrt{2}a=\lambda/\sqrt{2}$. The tight-binding parameters can be evaluated from the realistic lattice potential (\mbox{Appendix~\ref{app:wannier}}), yielding $t_1=-0.05$, $t_2=-0.07$, $t_3=-0.07$, $\varepsilon=-0.66$ in units of $E_{rec}$ (triangular lattice, $\eta_{sp}=0.19$) and $t_1 = t_2 = t_3 = 0.06$ in units of $E_{rec}$ (hexagonal lattice). 

\section{Derivation of the Hamiltonian}\label{app:wannier}

We start from Hamiltonian in co-moving frame,
\begin{equation}
    \hat{H}_{cm}(\tau)=\frac{\hat{p}^2}{2m}+V_{lat}(\hat{r})-\textbf{F}(\tau)\cdot\hat{r}~,
\end{equation}
where $\hat{p}$ and $\hat{r}$ are momentum and position operators, respectively, and $\textbf{F}=-m\Ddot{\textbf{r}}_m(\tau)=-\dot{\textbf{A}}(\tau)$. The Hamiltonian has two parts. The first two terms include the static lattice Hamiltonian and the last term represents a time-dependent dispersion.

Next, we expand $\hat{H}_{cm}(\tau)$ in Wannier basis~\cite{marzari_maximally_1997},
\begin{equation}
    w_{n,\textbf{R}}(\textbf{r}) \equiv w_{n}(\textbf{r}-\textbf{R})=\frac{V}{(2\pi)^2}\int_{BZ}e^{-i\textbf{k}\cdot\textbf{R}}\phi_{n,\textbf{k}}(\textbf{r})\,d\textbf{k}~,
\end{equation}
where $n$ is band index, $\phi_{n,\textbf{k}}(\textbf{r})=e^{i\textbf{k}\textbf{r}}u_{n,\textbf{k}}(\textbf{r})$ is Bloch state, $\textbf{R}=n_1 \textbf{a}_1+n_2  \textbf{a}_2$ is arbitrary lattice vector, and $\textbf{a}_1$, $\textbf{a}_2$ are primitive translation vectors in two dimensions.

In second quantization, we can rewrite the field operator in Wannier basis as
\begin{equation}
    \hat{\psi}(\textbf{r})=\sum_{n, \textbf{R}}w_{n,\textbf{R}}^*(\textbf{r})\hat{a}_{n,\textbf{R}}~,
\end{equation}
where $\hat{a}_{n,\textbf{R}}$ is the annihilation operator.

The first part of the co-moving Hamiltonian can be expanded as
\begin{equation}
    \hat{H}_0(\tau)=\sum_{n,m,\textbf{R},\textbf{R'}}\hat{a}_{n,\textbf{R}}^{\dagger}\hat{a}_{m,\textbf{R'}}\int w_{n,\textbf{R}}(\textbf{r})\left[\frac{\hat{p}^2}{2m}+V_{lat}(\hat{r})\right]w_{m,\textbf{R'}}^*(\textbf{r})\,d\textbf{r}~.
\end{equation}

Using the fact that the static Hamiltonian cannot mix different bands and the Wannier functions are exponentially localized, we only consider the onsite and nearest-neighbour couplings in individual bands,
\begin{equation}
    \begin{aligned}
        \hat{H}_0(\tau)=\sum_{n,\textbf{R}}\left\{\hat{a}_{n,\textbf{R}}^{\dagger}\hat{a}_{n,\textbf{R}}\varepsilon_n+\sum_{\textbf{b}_i}\left[\hat{a}_{n,\textbf{R}}^{\dagger}\hat{a}_{n,\textbf{R}+\textbf{b}_i}t_i+h.c.\right]\right\}~,
    \end{aligned}
\end{equation}
\begin{equation}
    \varepsilon_n=\int w_{n,\textbf{R}}(\textbf{r})\left[\frac{\hat{p}^2}{2m}+V_{lat}(\hat{r})\right]w_{n,\textbf{R}}^*(\textbf{r})\,d\textbf{r}~,
\end{equation}
\begin{equation}
    t_i=\int w_{n,\textbf{R}}(\textbf{r})\left[\frac{\hat{p}^2}{2m}+V_{lat}(\hat{r})\right]w_{n,\textbf{R}+\textbf{b}_i}^*(\textbf{r})\,d\textbf{r}~.
\end{equation}
$\textbf{b}_i=j\textbf{a}_1+k\textbf{a}_2$ are the nearest-neighbour tunneling vectors.
In the configuration of triangular and hexagonal lattices, $i\in\{1,2,3\}$. $\varepsilon_n$ is the band center energy and $t_i$ is the nearest-neighbour tunneling amplitude.

Now we consider the driving term which can couple different bands on-site but has negligible inter-site effects, i.e.
\begin{equation}
    \begin{aligned}
        \hat{H}_1(\tau)&=-\sum_{n,m,\textbf{R},\textbf{R'}}\hat{a}_{n,\textbf{R}}^{\dagger}\hat{a}_{m,\textbf{R'}}\int w_{n,\textbf{R}}(\textbf{r})\left[\textbf{F}(\tau)\cdot\hat{r}\right]w_{m,\textbf{R'}}^*(\textbf{r})\,d\textbf{r}
        \\&=-\sum_{n,\textbf{R}}\hat{a}_{n,\textbf{R}}^{\dagger}\hat{a}_{n,\textbf{R}}\int w_{n,\textbf{R}}(\textbf{r})\left[\textbf{F}(\tau)\cdot\hat{r}\right]w_{n,\textbf{R}}^*(\textbf{r})\,d\textbf{r}
        -\sum_{n,m\neq n,\textbf{R}}\hat{a}_{n,\textbf{R}}^{\dagger}\hat{a}_{m,\textbf{R}}\int w_{n,\textbf{R}}(\textbf{r})\left[\textbf{F}(\tau)\cdot\hat{r}\right]w_{m,\textbf{R}}^*(\textbf{r})\,d\textbf{r}
        \\&=-\sum_{n,\textbf{R}}\left[\hat{a}_{n,\textbf{R}}^{\dagger}\hat{a}_{n,\textbf{R}}\textbf{F}(\tau)\cdot\textbf{R}+\sum_{m\neq n}\hat{a}_{n,\textbf{R}}^{\dagger}\hat{a}_{m,\textbf{R}}a\textbf{F}(\tau)\cdot\eta_{nm}\hat{r'}\right]~.
    \end{aligned}
\end{equation}
Here we define the inter-band coupling as
\begin{equation}
    \eta_{nm}=\frac{1}{b}\left\lvert\int w_{n,\textbf{R}}(\textbf{r})\hat{r} w_{m,\textbf{R}}^*(\textbf{r})\,d\textbf{r}\right\rvert.
\end{equation}

By numerically evaluating $\eta_{sp}$ according to the method in \cite{bissbort_dynamical_2013, uehlinger_artificial_2013}, we justify that the dipole-like coupling term $\int w_{m,\textbf{R}}(\textbf{r})\hat{r} w_{n,\textbf{R}}^*(\textbf{r})\,d\textbf{r}$ between s and p band is along the direction $\hat{r'} = \hat{r} = \hat{e}_x + \hat{e}_y$.
In addition, we evaluate tunnelings and band center energies through calculating the Wannier functions of atoms in a realistic lattice potential.

The on-site driving term $\hat{a}_{n,\textbf{R}}^{\dagger}\hat{a}_{n,\textbf{R}}\textbf{F}(\tau)\cdot\textbf{R}$ breaks translation symmetry, so we need to rotate it away by applying the unitary
\begin{equation}
    \hat{U}(\tau)=\exp\left(-i\sum_{\textbf{R}n}\chi_{\textbf{R}(\tau)}\hat{a}_{n,\textbf{R}}^{\dagger}\hat{a}_{n,\textbf{R}}\right),
\end{equation}
\begin{equation}
    \hat{H}'(\tau)=\hat{U}^{\dagger}(\tau)\hat{H}(\tau)\hat{U}(\tau)-i\hbar\hat{U}^{\dagger}(\tau)\frac{\partial}{\partial \tau}\hat{U}(\tau),
\end{equation}
where $\chi_{\textbf{R}}(\tau)=\frac{m}{\hbar} \textbf{R}\cdot\dot{\textbf{r}}_m(\tau)$.
Then, transforming the Hamiltonian into quasi-momentum space and only considering the lowest two bands, we get the final Hamiltonian,
\begin{equation}
    \begin{aligned}
        \hat{H}_{\textbf{q}}(\tau)=\sum_{n \in \{s, p\}}\biggl\{\varepsilon_{n}\hat{a}_n^{\dagger}\hat{a}_n+\sum_{i=1}^{Z}\left[t_{n, i}\left(e^{i(\theta_i(\tau)+\textbf{q}\cdot\textbf{b}_i)}\hat{a}_n^{\dagger}\hat{a}_n+h.c.\right)\right]-\textbf{F}(\tau)\cdot\hat{r}\sum_{m \in \{s, p \} \neq n}\eta_{nm}\hat{a}_n^{\dagger}\hat{a}_{m}\biggl\},
    \end{aligned}
\end{equation}
with the time-periodic Peierls phase $\theta_i(\tau)=\chi_{\textbf{R}+\textbf{b}_i}(\tau)-\chi_{\textbf{R}}(\tau)=\frac{m}{\hbar} \textbf{b}_i\cdot\dot{\textbf{r}}_m(\tau)$.

\section{Analytical Hamiltonian by the high-frequency expansion}\label{app:HFE}

In the following, we write down the effective Floquet Hamiltonian, analytically derived using the high-frequency expansion (HFE) in tight-binding approximation.
The idea of the HFE is to decompose the original time-dependent Hamiltonian into Fourier components and obtain the effective Hamiltonians, then neglect the small high frequency terms. The effective Hamiltonian is decomposed into terms of different orders,
\begin{equation}
    \hat{H}_{\mathrm{eff}}=\sum_{n=0}^{\infty}\hat{H}_{\mathrm{eff}}^{(n)},
\end{equation}
which is in inverse power of $\omega$, i.e. $\hat{H}_{\mathrm{eff}}^{(n)}\sim\omega^{-n}$. If the frequency is much higher than other energy scales in the system, the high-order effective Hamiltonian can be truncated. Here we only use the first two orders~\cite{eckardt_high-frequency_2015}, 

\begin{equation}
    \hat{H}_{\mathrm{eff}}^{(0)}=\frac{1}{T}\int_{0}^{T}\hat{H}(\tau)\,d\tau\equiv \hat{H}_0~,
\end{equation}
\begin{equation}
    \hat{H}_{\mathrm{eff}}^{(1)}=\frac{1}{\hbar\omega}\sum_{l=1}^{\infty}\frac{1}{l}\left[\hat{H}_l, \hat{H}_{-l}\right] = \sum_{l=1}^{\infty}\hat{H}_{\mathrm{eff}, l}^{(1)}~.
\end{equation}
Here, $\hat{H}_l$ are the Fourier components of the time-dependent Hamiltonian, $\hat{H}(\tau)=\sum_{l=-\infty}^{\infty}\hat{H}_le^{il\omega\tau}$.
\begin{table}[htbp]
    \centering
    \setlength\tabcolsep{3pt}
    \begin{tabular}{c c}
    \hline\\[-3pt]
    \multirow{3}{*}{$\hat{H}_{\mathrm{eff}}^{(0)}$} & $\sigma_x\eta_{sp}\hbar\omega[K_x+K_y\cos(\varphi)]/2$ \\[6pt]
    & $\sigma_y\eta_{sp}\hbar\omega K_y\sin(\varphi)/2$ \\[6pt]
    & \makecell[c]{$\sigma_z\{\varepsilon+\hbar\omega/2+t_1\cos[(q_x-q_y)b]J_0(2K_x)+t_2\cos(q_x b)[J_0(K_x)J_0(K_y)-2J_{1}(K_x)J_{1}(K_y)\cos(\varphi)+$\\$2J_{2}(K_x)J_{2}(K_y)\cos(2\varphi)]+t_3\cos(q_y b)[J_0(K_x)J_0(K_y)+2J_{1}(K_x)J_{1}(K_y)\cos(\varphi)+2J_{2}(K_x)J_{2}(K_y)\cos(2\varphi)]\}$} \\[12pt]
    \hline\\[-3pt]
    \multirow{3}{*}{$\hat{H}_{\mathrm{eff}, 2}^{(1)}$} & \makecell[c]{$\sigma_x\eta_{sp}\{-t_1\cos[(q_x-q_y)b]J_2(2K_x)[K_x+K_y\cos(\varphi)]-t_2\cos(q_x b)\{J_1(K_x)J_1(K_y)[K_y+K_x\cos(\varphi)]+$\\$
    J_0(K_y)J_2(K_x)[K_x+K_y\cos(\varphi)]+J_0(K_x)J_2(K_y)[K_y\cos(\varphi)
    +K_x\cos(2\varphi)]\}+t_3\cos(q_y b)\{J_1(K_x)J_1(K_y)$\\$[K_y+K_x\cos(\varphi)]-J_0(K_y)J_2(K_x)[K_x+K_y\cos(\varphi)]-J_0(K_x)J_2(K_y)[K_y\cos(\varphi)+K_x\cos(2\varphi)]\}\}/2$} \\[18pt]
    & \makecell[c]{$\sigma_y\eta_{sp}\sin(\varphi)\{t_1\cos[(q_x-q_y)b]J_2(2K_x)K_y-t_2\cos(q_x b)\{J_1(K_x)J_1(K_y)K_x-
    J_0(K_y)J_2(K_x)K_y+J_0(K_x)J_2(K_y)$\\$[K_y
    +2K_x\cos(\varphi)]\}+t_3\cos(q_y b)\{J_1(K_x)J_1(K_y)K_x+J_0(K_y)J_2(K_x)K_y-J_0(K_x)J_2(K_y)[K_y+2K_x\cos(\varphi)]\}\}/2$} \\[12pt]
    & $\sigma_z\eta^2\hbar\omega[K_x^2+K_y^2+2K_x K_y\cos(\varphi)]/8$ \\[6pt]
    \hline
    \end{tabular}
    \caption{\label{tab:table-tri_h1}Effective Hamiltonian for the triangular lattice under single frequency driving resonant with the gap}
\end{table}

Table~\ref{tab:table-tri_h1} is the effective Hamiltonian for the triangular lattice when the driving field is resonant with the gap energy. The constant terms are much larger than the quasimomentum dependent terms in off-diagonal, which means the on-site inter-band couplings are dominant.
We safely neglect the Bessel functions higher than the second order as the driving strength $K \leq 1$ remains in weak driving regime.

Table~\ref{tab:table-tri_h2} shows the case of single-frequency driving resonant with the half the gap. Now the off-diagonal terms are dominantly quasimomentum dependent. The Hamiltonian for the triangular lattice in two-frequency driving scheme are not listed here.
\begin{table}[htbp]
    \centering
    \setlength\tabcolsep{3pt}
    \begin{tabular}{c c}
    \hline\\[-3pt]
    $\hat{H}_{\mathrm{eff}}^{(0)}$ & \makecell[c]{$\sigma_z \{\varepsilon+\hbar\omega+t_1\cos[(q_x-q_y)b]J_0(2K_x)+t_2\cos(q_x b)[J_0(K_x)J_0(K_y)-2J_1(K_x)J_1(K_y)\cos(\varphi)+$\\$2J_2(K_x)J_2(K_y)\cos(2\varphi)]+t_3\cos(q_y b)[J_0(K_x)J_0(K_y)+2J_1(K_x)J_1(K_y)\cos(\varphi)+2J_2(K_x)J_2(K_y)\cos(2\varphi)]\}$} \\[12pt]
    \hline\\[-3pt]
    \multirow{3}{*}{$\hat{H}_{\mathrm{eff}, 1}^{(1)}$} & \makecell[c]{$\sigma_x\eta_{sp}\{-t_1 \sin[(q_x-q_y)b]J_1(2K_x)K_y\sin (\varphi)-t_2\sin(q_x b)\sin(\varphi)\{K_y J_0(K_y) J_1(K_x) + K_x J_1(K_y) J_2(K_x) -$\\$ J_1(K_x) J_2(K_y) [K_y + 2K_x \cos(\varphi) +2K_y \cos(2\varphi)] + J_0(K_x) J_1(K_y) [K_x + 2K_y \cos(\varphi)]\} $\\$- t_3\sin(q_y b)\sin(2\varphi)\{K_y J_0(K_x) J_1(K_y) + J_1(K_x) J_2(K_y) [K_x + 2K_y \cos(\varphi)]\}\}$} \\[18pt]
    & \makecell[c]{$\sigma_y\eta_{sp}\{t_1 \sin[(q_x-q_y)b]J_1(2K_x)[K_x + K_y\cos (\varphi)] + t_2\sin(q_x b)\{J_0(K_y) J_1(K_x) [K_x+ K_y \cos(\varphi)]- $\\$J_1(K_y) J_2(K_x) [K_y + K_x\cos(\varphi)] + J_1(K_y) J_0(K_x) [K_x \cos(\varphi) + K_y \cos(2\varphi)] - J_1(K_x) J_2(K_y)$\\$ [K_x \cos(2\varphi) + K_y \cos(3\varphi)]\} +  t_3\sin(q_y b)\sin(2\varphi)\{- J_0(K_y) J_1(K_x)[K_x + K_y\cos(\varphi)] - J_1(K_y) J_2(K_x)$\\$ [K_y + K_x \cos(\varphi)]+ J_1(K_y) J_0(K_x) [K_x \cos(\varphi) + K_y \cos(2\varphi)] + J_1(K_x) J_2(K_y) [K_x \cos(2\varphi) + K_y\cos(3\varphi)]\}\}$} \\[24pt]
    & $\sigma_z\eta_{sp}^2\hbar\omega[K_x^2+K_y^2+2K_x K_y\cos(\varphi)]/4$ \\[6pt]
    \hline \\[-3pt]
    \multirow{3}{*}{$\hat{H}_{\mathrm{eff}, 3}^{(1)}$} & \makecell[c]{$\sigma_x\eta_{sp}\{-t_2\sin(q_x b)\{K_x J_1(K_y) J_2(K_x) + J_1(K_x) J_2(K_y) [K_y + 2K_x \cos(\varphi)]\}$\\$ + t_3\sin(q_y b)\{-K_x J_1(K_y) J_2(K_x) +J_1(K_x) J_2(K_y) [K_y + 2K_x \cos(\varphi)]\}\}/3$} \\[12pt]
    & \makecell[c]{$\sigma_y\eta_{sp}\{t_2\sin(q_x b)\{J_1(K_y) J_2(K_x)[K_y + K_x\cos(\varphi)] + J_1(K_x) J_2(K_y) [K_y \cos(\varphi) + K_x \cos(2\varphi)]\}$\\$ + t_3\sin(q_y b)\{J_1(K_y) J_2(K_x)[K_y + K_x \cos(\varphi)] - J_1(K_x) J_2(K_y) [K_y\cos(\varphi) + K_x \cos(2\varphi)]\}\}/3$} \\[12pt]
    & $\sigma_z\eta_{sp}^2\hbar\omega[K_x^2+K_y^2+2K_xK_y\cos(\varphi)]/12$ \\[6pt]
    \hline
    \end{tabular}
    \caption{\label{tab:table-tri_h2}Effective Hamiltonian for the triangular lattice under single frequency driving resonant with the half the gap}
\end{table}
\begin{table}
    \centering
    \setlength\tabcolsep{3pt}
    \begin{tabular}{c c}
    \hline \\[-3pt]
    \multirow{2}{*}{$\hat{H}_{\mathrm{eff}}^{(0)}$} &  $\sigma_x\{[t_1\cos(q_y b) + t_3\cos(q_x b)] J_0(K_{y1}) + t_2\cos[(q_x-q_y)b]J_0(K_{x1})\}$\\[6pt]
     & $\sigma_y\{[t_1\sin(q_y b) - t_3\sin(q_x b)] J_0(K_{y1}) - t_2\sin[(q_x-q_y)b]J_0(K_{x1})\}$\\[6pt]
    \hline \\[-3pt]
    $\hat{H}_{\mathrm{eff}}^{(1)}$ & \makecell[c]{$-\sigma_z\sum_{n=1}^{\infty}\frac{4}{n\hbar\omega} t_2 J_n(K_{x1}) J_n(K_{y1}) \left[(-1)^{n+1}t_1 \sin(q_x b) +t_3 \sin(q_y b)\right]\sin(n\varphi_{y1})$}\\[6pt]
    \hline
    \end{tabular}
    \caption{\label{tab:table-hex_h1}Effective Hamiltonian for the hexagonal lattice under single frequency off-resonant driving}
\end{table}

For the hexagonal lattice, Table~\ref{tab:table-hex_h1} gives the general Hamiltonian under single-frequency off-resonant driving, while Table~\ref{tab:table-hex_h2} is the Hamiltonian under two-frequency off-resonant driving. Note that the second order Bessel functions are omitted in $\hat{H}_{\mathrm{eff}}^{(1)}$ in Table~\ref{tab:table-hex_h2} but kept in the numerical calculations.
\begin{table}[htbp]
    \centering
    \setlength\tabcolsep{3pt}
    \begin{tabular}{c c}
    \hline \\[-3pt]
    \multirow{2}{*}{$\hat{H}_{\mathrm{eff}}^{(0)}$} &  \makecell[c]{$\sigma_x\{[t_1\cos(q_y b) + t_3\cos(q_x b)] J_0(K_{y1}) J_0(K_{y2}) + t_2\cos[(q_x-q_y)b]J_0(K_{x1}) J_0(K_{x2}) - $\\$2[t_1 \sin(q_x b) + t_3\sin(q_y b)] J_1(K_{y2}) J_2(K_{y1}) \sin(2\varphi_{y1} - \varphi_{y2}) + 2 t_2 \sin[(q_x - q_y)b] J_1(K_{x2}) J_2(K_{x1}) \sin(\varphi_{x2})\}$}\\[12pt]
     & \makecell[c]{$\sigma_y\{[t_1\sin(q_y b) - t_3\sin(q_x b)] J_0(K_{y1}) J_0(K_{y2}) - t_2\sin[(q_x-q_y)b]J_0(K_{x1}) J_0(K_{x2}) + $\\$2[t_1 \cos(q_x b) - t_3\cos(q_y b)] J_1(K_{y2}) J_2(K_{y1}) \sin(2\varphi_{y1} - \varphi_{y2}) + 2 t_2 \cos[(q_x - q_y)b] J_1(K_{x2}) J_2(K_{x1}) \sin(\varphi_{x2})\}$}\\[12pt]
    \hline \\[-3pt]
    $\hat{H}_{\mathrm{eff}}^{(1)}$ & \makecell[c]{$\sigma_z \frac{2}{3\hbar\omega} \{6J_0(K_{y2}) J_1(K_{y1}) [t_2 J_1(K_{x1}) J_1(K_{x2}) (t_1\cos(q_x b) - t_3 \cos(q_y b))\cos(\varphi_{x2} - \varphi_{y1}) + (t_1^2 - t_3^2)$\\$ J_1(K_{y1}) J_1(K_{y2}) \cos(2\varphi_{y1} - \varphi_{y2})] - 6t_2 J_0(K_{x2}) J_1(K_{x1}) [t_2 J_1(K_{x1}) J_1(K_{x2}) \cos(\varphi_{x2})+ J_1(K_{y1}) J_1(K_{y2})$\\$(t_1 \cos(q_x b) + t3 \cos(q_y b)) \cos(\varphi_{y1} - \varphi_{y2}) - J_1(K_{y1}) J_0(K_{y2})(t_1 \sin(q_x b) + t3 \sin(q_y b)) \sin(\varphi_{y1})]$\\$ - t_2 J_1(K_{x2}) J_1(K_{y2}) [ 3 J_0(K_{x1}) J_0(K_{y1}) (t_1 \sin(q_x b) + t_3 \sin(q_y b))\sin(\varphi_{x2} - \varphi_{y2}) - 2J_1(K_{x1}) J_1(K_{y1})$\\$ (t_1 \sin(q_x b) - t_3\sin(q_y b))(\sin(\varphi_{x2} - \varphi_{y1} - \varphi_{y2}) + 3 \sin(\varphi_{x2} + \varphi_{y1} - \varphi_{y2}))]\}$}\\[30pt]
    \hline
    \end{tabular}
    \caption{\label{tab:table-hex_h2}Effective Hamiltonian for the hexagonal lattice under two frequency off-resonant driving}
\end{table}

\end{document}